\documentclass[aps,pra,groupedaddress,twocolumn,showpacs, showkeys]{revtex4}
\usepackage{graphicx}
\usepackage{amssymb}
\usepackage{subfigure}
\usepackage{amsmath}
\usepackage{float}
\usepackage{amsfonts}
\usepackage{bm}
\usepackage{color}
\usepackage{dsfont}
\newcommand{\beq}{\begin{equation}}
\newcommand{\eeq}{\end{equation}}

\newcommand{\ket}[1]{\ensuremath{\left|{#1}\right\rangle}}
\newcommand{\bra}[1]{\ensuremath{\left\langle{#1}\right|}}

\begin{document}

	\preprint{}

	\title{Implementing positive-operator-valued-measurement elements in photonic circuits for performing minimum quantum state tomography of path qudits} 

	\author{W. R. Cardoso}
	\email{will.rodrigues.fis@gmail.com}
	\author{D. F. Barros}
	\author{M. R. Barros}
	\author{S. P\'adua}
	\affiliation{Departamento de F\'{\i}sica, Universidade Federal de Minas Gerais, 31270-901 Belo Horizonte, Minas Gerais, Brazil}

	\begin{abstract}
    Manipulation of qudits in optical tables is a difficult and non-scalable task. The use of integrated optical circuits opens new possibilities for the generation, manipulation and characterization of high dimensional states besides the ease of transmission of these states through an optical fiber. In this work, we propose photonic circuits to perform minimum quantum state tomography of path qudits and show how to determine all the constituents parameters of these circuits (beam splitters and phase shifters). Our strategies were based on the symmetries of the involved POVMs suggested for minimum tomography and allowed us to obtain interferometers smaller than those obtained by other already known methods. The calculations of the transmittances and reflectivities of the beam splitters were made using the definition of probability operators in an extended Hilbert spaces and the application of the Naimark's theorem. The employment of equidistant states for the definition of the POVM elements allowed us to develop a recipe applicable to the tomography of qudits of any dimension, generalizing our scheme.
	\end{abstract}

	\pacs{42.50.Ex, 03.65.Wj, 42.82.Et, 07.60.Ly}
	\keywords{quantum tomography, quantum information, optical waveguides, photonic circuits, qudits}

	\maketitle
	\section{Introduction}
	A qubit corresponds to the fundamental unit of quantum information in the same sense as the bit corresponds to the fundamental classical unit of information. When we are referring to quantum states with a Hilbert space with dimension $N > 2$, then we speak of \textit{qudits} \cite{rungta2001qudit, thew2004bell, de2002creating, thew2002qudit, neves2005generation, moreva2006realization, li2008generation}. The use of qudits opens new prospects for quantum information processing, such as the enhancement of security for quantum key distribution, the increase of channel capacity, higher noise resistance and quantum computing \cite{Ursin17, Torres12, kues2017chip, wang2018multidimensional, walborn2006quantum, wang2017qudit, wang2018proof, etcheverry2013quantum, gedik2015computational, niu2018qudit, chau2015quantum, islam2017provably, pal2018entanglement}.

	Manipulating states of qudits in interferometers at optical tables is a non-scalable and challenging task. Interferometers have been set for characterizing quantum states \cite{agnew2011tomography,Pimenta13,bent2015experimental,ikuta2017implementation,martinez2019experimental}, controlling, detecting or measuring entanglement of qudit systems \cite{kues2017chip,wang2018multidimensional, gutierrez2012experimental, gutierrez2014detection} or simulating noisy channels \cite{marques2015experimental,varga2018characterizing}. The use of photonic circuits written in glasses by femtosecond direct laser writing technique opens new possibilities for the generation, manipulation, and characterization of high dimensional states, besides the direct transmission of these states through an optical fiber \cite{Politi08,Marshall09,shadbolt2012generating,o2009photonic,caspani2017integrated,smith2009phase,Davis96,Streltsov01}. Using longitudinal interferometers, we can implement quantum operations in two qudits states, including transforming them from antisymmetric states to symmetric ones, making possible to construct interferometers for simulation of open systems and their characterization \cite{Walther12, Koch12,harris2017quantum,kerman2017multiloop}. Remote preparation of arbitrary qudits for their use in quantum communication protocols is another possible application for these circuits \cite{Fan08, Yin11,wang2015controlled,cao2016flexible}.

    Recently, the generation and analysis of a two qutrits system in a silicon photonic circuit generated with the lithographic technique was demonstrated \cite{Schaeff15}. In this domain, we can use such result to produce states of two qudits from an integrated photonic chip drawn on glass, after they receive photon pairs generated by spontaneous parametric down-conversion (SPDC). We aim here the construction of photonic circuits that, by implementing POVM  (Positive Operator-Valued Measure) elements, are able to perform minimum full quantum state tomography of qudits, defined as an experimental technique that allows the reconstruction of the density matrix of a quantum system in a unknown state with a minimum number of measurements \cite{Raymer94, Home2006, Riebe2006,six2016quantum,delaney2019quantum,xin2017quantum}. These techniques satisfy a demand in current computer and telecommunications technologies: the ability to execute quantum information protocols on small devices that integrate a larger hardware design. Our accomplishment is a contribution in the development of quantum information tasks in chips. It contributes to the technology needed to reach compact and portable devices for quantum protocols.
    
    Quantum state tomography on chip has been a subject of interest to various research groups. It was demonstrated, theoretically and experimentally, the accomplishment of quantum tomography in photonic chips through quantum walks \cite{Titchener16,titchener2018scalable}. The authors used in their experiment a circuit with N inputs and M outputs to characterize a photon-number state. The detection of the photons of all the outputs is done simultaneously, optimizing the experiment duration. Since it is not necessary to reconfigure the experimental apparatus, the occurrence of errors is also reduced. Circuits that do not require reconfiguration are called static circuits \cite{oren2017quantum}. 
    
    In this work, we present an alternative form of performing full quantum tomography on photonic chips, by the implementation of POVM elements in static circuits, whose detection at all outputs is made simultaneously. All necessary POVM elements are implemented at the same time by the circuit and measured at the different photonic circuit exits simultaneously. All necessary measurement probabilities are obtained from the photon counts at the circuit outputs in one measurement time interval. This is important because it minimizes the noise introduced by the experimental apparatus during the detection time. The proposed tomography method uses SIC-POVM (Symmetric Informationally Complete Positive Operator-Valued Measure) elements, allowing a full quantum state tomography with the minimum number of measurements and with a symmetric POVM elements distribution in Hilbert space that optimizes the system density operator reconstruction \cite{Pimenta13,PaivaSanchez,petz2012efficient,petz2012optimal,rastegin2014notes,renes2004symmetric}.
    
    Our proposal is based on the application of the Naimark's theorem. The POVM can't be realized by an unitary transformation followed by a projective measurement in the original qudit space. Therefore, the initial qudit Hilbert space is expanded and the realization of the POVM elements occurs by applying unitary transformations followed by projection measurements in this extended space  \cite{Barnett,beneduci2010infinite,coecke2008povms,beneduci2010unsharpness,tabia2012experimental}. This task can be realized in multiport interferometers, composed of beam splitters and phase shifters, built by using, for example, the Reck's proposition  \cite{Reck94}. An important advance was done by Clements \textit{et al.}, that propose a different design for the multiport interferometer that, by using the same number of components, decreases the optical depth of the interferometer \cite{clements2016optimal}. This quantity is important for fabricating smaller interferometers with smaller losses. 
    
    The unitary transformations in the extended Hilbert space in our circuit design uses a smaller number of beam splitters than Reck \textit{et al.} and Clements \textit{et al.} general designs with a smaller optical depth, what is a clear advantage for this specific task. It is necessary to mention that the proposed circuit design realizes full quantum tomography with the minimum number of measurements and no hypothesis about the initial system state are necessary. Others schemes are able to realize quantum tomography with a number of measurements smaller than the minimum for a particular class of systems states using prior knowledge of the input state \cite{oren2017quantum}.
    
    Finally, the organization of the rest of the paper is as follows: in Sec.~\ref{sec:tomography}, we briefly discuss some useful concepts about minimum state tomography for qubits. After this short introduction, we present in Sec.~\ref{sec:circuitsqubits} the proposal for construction of the tomographical circuits for this case. We then proceed in Sec.~\ref{sec:circuitsqutrit} with the circuit's proposal in the qutrit case. In Sec.~\ref{sec:circuitsNd}, we generalize our proposal for $N$ dimensions. Finally, we conclude in Sec.~\ref{sec:conclusao}.

	\section{Minimum quantum state tomography for qubits}
	\label{sec:tomography}

    As already mentioned in the introduction, quantum state tomography is an experimental technique that allows the reconstruction of the density matrix of an unknown quantum system state. Its realization consists in the production of a large number of identically prepared system states together with a series of measurements of the physical quantities that it describes. When the tomography is made using the least possible number of operations, it is called a minimum full quantum state tomography \cite{Rehacek, Pimenta10}. To perform a minimal full tomography on a qubit system, one needs a POVM with four different elements $\hat{E}_{j}$ to be implemented in the initial state $\hat{\rho}$. The outcome's counting rate $c_j$ is
     \begin{equation}
	    c_{j}=AP_j,
	    \label{count}
	\end{equation}
	where $A$ is a constant proportional to the detector efficiency and the number of identically prepared states in one second and $P_j=\text{Tr}[\hat{\rho}\hat{E}_{j}]$ is the probability of detecting one of the output states after $E_j$ is implemented. By implementing these POVM elements, it is possible to determine the qubit state since the number of counts is going to be accessible.

	As seen in \cite{Pimenta10}, the necessary POVM elements to be implemented are:
	\begin{equation} \label{POVMs}
		\hat{E}_{k}=\dfrac{1}{2}\ket{\varphi_{k}}\bra{\varphi_{k}}
	\end{equation}
	with $k=1,2,3,4$ and
	\begin{subequations}
		\begin{align}
		\ket{\varphi_1} & = \omega \ket{0} + i \upsilon \ket{1}, \\
		\ket{\varphi_2} & = \omega \ket{0} - i \upsilon \ket{1}, \\
		\ket{\varphi_3} & = \upsilon \ket{0} - \omega \ket{1}, \\
		\ket{\varphi_4} & = \upsilon \ket{0} + \omega \ket{1},
		\end{align}
	\end{subequations}	
	where $\omega=\sqrt{2/3}$ and $\upsilon=\sqrt{1/3}$. If the density operator $\hat{\rho}$ is written as $\hat{\rho}=(\hat{I}+\vec{r}\cdot\vec{\sigma})/2$, where $\vec{r}=(r_x, r_y, r_z)$ and $\vec{\sigma}=(\hat{\sigma}_x,\hat{\sigma}_y,\hat{\sigma}_z)$ is the Pauli vector, we need to determine the values of the quantities $r_x$, $r_y$ and $r_z$ from the experimental outcome's counting rates $c_j$ (Eq.~\eqref{count}) in order to reconstruct the initial state density matrix $\hat{\rho}$. By using the POVM elements shown in Eq.~\eqref{POVMs} in Eq.~\eqref{count}, we obtain
	\begin{subequations}
		\begin{align}
			r_x=\dfrac{3(c_4-c_3)}{A\sqrt{2}},\qquad \\
		r_y=\dfrac{3(c_1-c_2)}{A\sqrt{2}}, \qquad	\\	
		r_z=\dfrac{3(c_1+c_2-c_3-c_4)}{A},
		\end{align}
	\end{subequations}	
	where $A=c_1+c_2+c_3+c_4$.
	\section{Construction of the tomographical circuits for qubits}
	\label{sec:circuitsqubits}

    Any unitary operator of finite size can be constructed in the laboratory using an appropriate combination of two optical devices: beam splitters (BS) and phase shifters \cite{Reck94}. Therefore, we can express an $N$-dimensional operator $\hat{U}'$ as:
    \begin{equation}\label{circuitomatrix}
	   \hat{U}'= \hat{T}_n \cdot \hat{T}_{n-1} \cdots \hat{T}_2 \cdot \hat{T}_1,
	\end{equation}
    where $\hat{T}_j$ is an $N$-dimensional matrix that represents the $j$th operation performed on the input state and $n$ is the number of operations required for the implementation of the matrix $\hat{U}'$. The $\hat{T}_j$ matrix is a block-diagonal matrix, where each block is composed from the elements of the BS matrix, defined as follows \cite{Reck94}:
    \begin{equation}
	    \hat{U}_{BS} =
	    \left[
	    \begin{array}{cc}
	        e^{i\phi}\sqrt{r}  & e^{i\phi}\sqrt{t} \\
	        \sqrt{t}           & -\sqrt{r}
	    \end{array}
	    \right],
	\end{equation}
    in which $t$ is the transmittance of the beam splitter and $r$ is its reflectivity. These parameters satisfy the equation $r+t=1$. The parameter $\phi$ indicates the phase difference between the outputs of the beam splitter. These are the building blocks we will use to create a scheme for minimal quantum state tomography in integrated chips.
    
\begin{figure}[]
    	\centering
    	\hspace*{-0.2cm} \includegraphics[height=4cm]{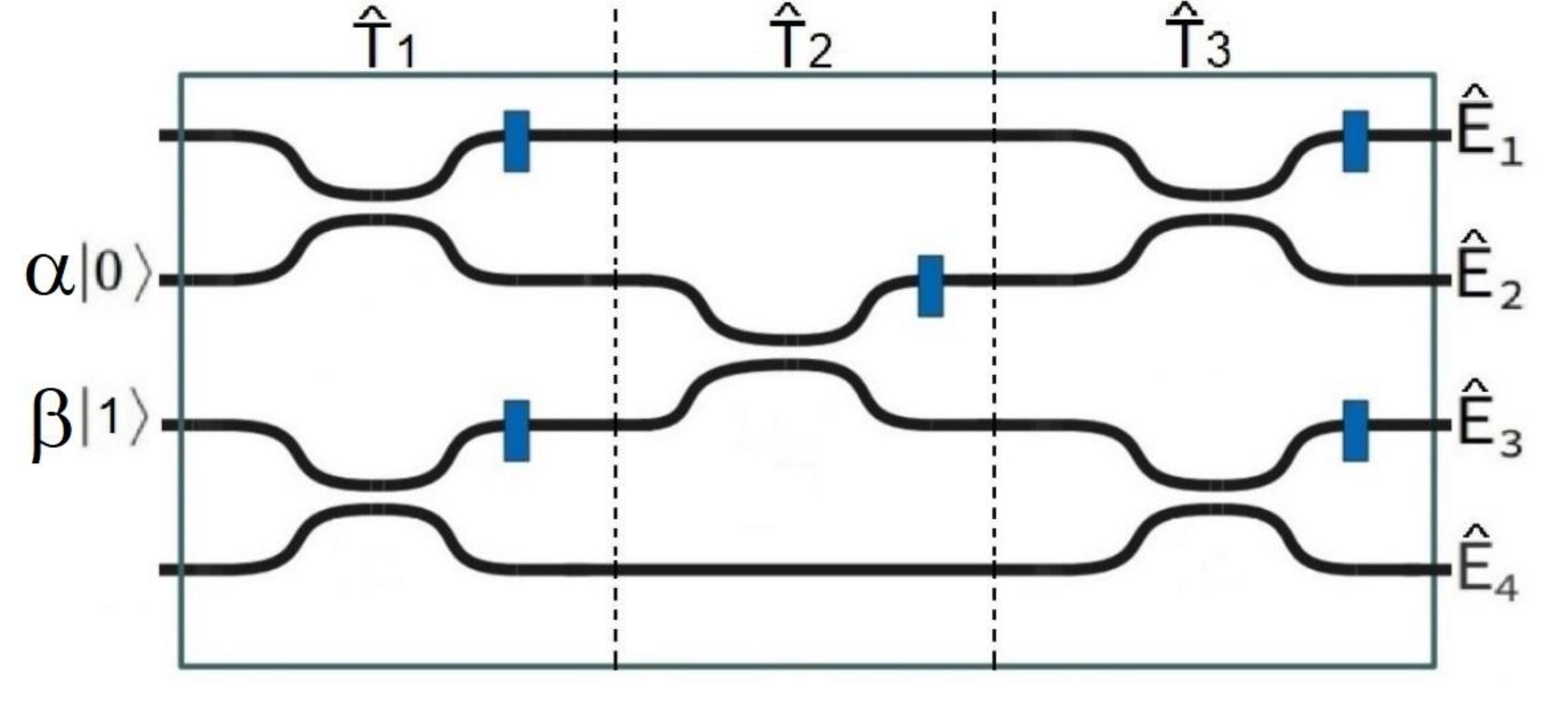}
    	\caption{Circuit proposal for the realization of the minimal state tomography on an one-qubit system. The black lines are the photon paths realized by the waveguides; the BS's are represented by the approximation of these lines and the phase shifters are the blue rectangles. This proposal was made on the assumption that the modes of the photonic qubit labelled as $\ket{0}$ and $\ket{1}$ are coupled to the second and third waveguides, respectively.}
    	\label{saidaspovm} 
    \end{figure}

    The first goal of this article is to propose a photonic circuit that implements the minimal full path state tomography on an one-qubit system. The simpler choice for the interferometer is shown in Fig.~\ref{saidaspovm}. In this configuration, each outcome of the interferometer corresponds to the implementation of one of the different maps $\hat{E}_k$, with $k=1,2,3,4$ (Eq.~\eqref{POVMs}), allowing us to obtain the outcomes simultaneously. The $\hat{T}_j$ operators that it constitutes (Fig.~\ref{saidaspovm}) are:   
	\begin{equation}
	    \hat{T}_1 =
	    \left[
	    \begin{array}{cccc}
	        e^{i \phi_{1}} \sqrt{r_1} & e^{i \phi_{1}} \sqrt{t_{1}} & 0                          & 0 \\
	        \sqrt{t_{1}}                & -\sqrt{r_{1}}               & 0                          & 0 \\
	        0                           & 0                           & e^{i\phi_{2}} \sqrt{r_{2}} & e^{i\phi_{2}} \sqrt{t_{2}} \\
	        0                           & 0                           & \sqrt{t_{2}}               & -\sqrt{r_{2}}
	    \end{array}
	    \right],
	\end{equation}
	\begin{equation}
	    \hat{T}_2 =
	    \left[
	    \begin{array}{cccc}
	        1 & 0                           & 0                           & 0 \\
	        0 & e^{i \phi_{3}} \sqrt{r_{3}} & e^{i \phi_{3}} \sqrt{t_{3}} & 0 \\
	        0 & \sqrt{t_{3}}                & -\sqrt{r_{3}}               & 0 \\
	        0 & 0                           & 0                           & 1
	    \end{array}
	    \right],
	\end{equation}
	\begin{equation}
	\hat{T}_3 =
	\left[
	\begin{array}{cccc}
	e^{i \phi_{4}} \sqrt{r_4} & e^{i \phi_{4}} \sqrt{t_{4}} & 0                          & 0 \\
	\sqrt{t_{4}}                & -\sqrt{r_{4}}               & 0                          & 0 \\
	0                           & 0                           & e^{i\phi_{5}} \sqrt{r_{5}} & e^{i\phi_{5}} \sqrt{t_{5}} \\
	0                           & 0                           & \sqrt{t_{5}}               & -\sqrt{r_{5}}
	\end{array}
	\right].
	\end{equation}
    
    Since we need to expand the Hilbert space, we will use as initial state $\ket{\psi}_0=\ket{\psi}\otimes\ket{0}$. The first state vector can be interpreted as a qubit system in the state $\ket{\psi}$, and the second as an \textit{ancilla} in the state $\ket{0}$. But, note that the state $\ket{\psi}_0=(\alpha,0,\beta,0)^T$ is different from the initial state shown in the Fig.~\ref{saidaspovm}, given by $\ket{\psi}_{in}=(0,\alpha,\beta,0)^T$. The input state $\ket{\psi}_{in}$ can be written in terms of the state $\ket{\psi}_0$, with no modifications in the circuit nor any relabeling, by using the transformation described by $\hat{M}$, defined by the following relation
    \begin{equation}
    \left[
    \begin{array}{cccc}
    0 \\ \alpha \\ \beta \\  0 
    \end{array}
    \right]=\hat{M}\cdot
    \left[
    \begin{array}{cccc}
    \alpha \\ 0 \\ \beta \\  0 
    \end{array}
    \right]
    \quad\Rightarrow\quad    
    \label{M}
	    \hat{M} =
	    \left[
	    \begin{array}{cccc}
		    0 & 1 & 0 & 0 \\
		    1 & 0 & 0 & 0 \\
		    0 & 0 & 1 & 0 \\
		    0 & 0 & 0 & 1
	    \end{array}
	    \right].
    \end{equation}
    
    It is worth noting that the transformation $\hat{M}$ is not implemented in practice. It is only a mathematical transformation. Thus, as our proposal supposes the input state as being  $\ket{\psi}_{in}$ and the theoretical calculation is made based on the initial state $\ket{\psi}_0$, we have
   \begin{equation}
   \hat{U}'\ket{\psi}_{in} = \hat{T}_3 \cdot \hat{T}_2 \cdot \hat{T}_1 \cdot \hat{M}\ket{\psi}_0.
   \end{equation} 
       
   Therefore, the operator that effectively would act on the state $\ket{\psi}_0$ is given by
    \begin{equation}\label{portaU}
	    \hat{U}_{qubit} = \hat{T}_3\cdot\hat{T}_2\cdot\hat{T}_1\cdot\hat{M}.
	\end{equation}
	
	The operator defined in Eq.~\eqref{portaU} is used in the process of determination of the circuit parameters, as shown in the Appendix A.

	The next task is to find the parameters $r_{j}$, $t_{j}$ and $\phi_{j}$ that compose $\hat{U}_{qubit}$ to complete the circuit description. 
	We start by examining the set of probability operators $\hat{\Pi}_{ml}$ acting on the system-ancilla Hilbert space. Such operators can be understood as the POVM elements associated with the detection of the photon at each circuit output. The action results of this POVM elements correspond to the projections over the states of the set $\{\ket{ml}\}$, with $m,l=0,1$, and each circuit output corresponds to one of the states $\ket{ml}$, i.e., $\ket{00}$, $\ket{01}$, $\ket{10}$ or $\ket{11}$. Here, the states $\ket{m}$ and $\ket{l}$ form a basis on each of the subspaces: system and \textit{ancilla}, respectively. The probability of detecting a photon, initially in the state described by the density operator $\hat{\rho}$, in one given output of the interferometer is given by
	\begin{equation}\label{probml}
		P_{ml}=\text{Tr}\left[\left(\hat{\rho}\otimes\ket{0}\bra{0}\right)\hat{\Pi}_{ml}\right],
	\end{equation}
    where $\ket{0}$ is the \textit{ancilla} state. This expression can also be written in terms of a POVM acting only in the subspace of the input qubit
	\begin{equation}
	P_{ml} = \text{Tr} \left[ \hat{\rho} \hat{E}_{ml} \right],
	\end{equation}
	where we can define the local POVM elements $\hat{E}_{ml}$ as
	\begin{equation} \label{traco2}
	\hat{E}_{ml} = \text{Tr}_2 \left[ \left( \hat{\mathbb{I}} \otimes \ket{0}\bra{0} \right) \hat{U}_{qubit}^\dagger\ket{ml}\bra{ml} \hat{U}_{qubit} \right]
	\end{equation}
    and the matrix elements of $\hat{E}_{ml}$ will be given by
    \begin{equation} \label{barneteq}
    \bra{p}\hat{E}_{ml}\ket{q} = \bra{p0} \hat{U}_{qubit}^{\dagger} \ket{ml} \bra{ml} \hat{U}_{qubit} \ket{q0}.
    \end{equation}    
    
	Each state $\ket{ml}$, in binary notation, corresponds to one outcome shown in Fig.~\ref{saidaspovm}, which allows us to make the exchange $\hat{E}_{ml}\rightarrow\hat{E}_k$, being $k$ the number that identifies each output. Based on this, the operators $\hat{E}_{k}$ in Eq.~\eqref{traco2} will be identified as the POVM element presented in Eq.~\eqref{POVMs}. From these relations, it is possible to determine the suitable circuit parameters.
	As shown in the Appendix A, the parameters necessary for implementing the minimal tomography, in this case, are:
	\begin{subequations}
		\begin{align}
		r_1 & = 1/3, & t_1 & = 2/3, & \phi_1 & = \pi/2, \\
		r_2 & = 1/3, & t_2 & = 2/3,  & \phi_2 & = 0, \\
		\label{swap}
		r_3 & = 0,  & t_3 & = 1, & \phi_3 & = 0, \\
		r_4 & = 1/2, & t_4 & = 1/2, & \phi_4 & = 0, \\
		r_5 & = 1/2, & t_5 & = 1/2, & \phi_5 & = 0.
		\end{align}
	\end{subequations}

	By manufacturing a photonic circuit (Fig.~\ref{saidaspovm}) with the parameters presented in Eq. (17), we obtain a specific interferometer with five beam splitters and one phase shifter. This circuit implements POVM by doing unitary transformations and projective measurements in a extended Hilbert space with less optical elements and a smaller optical depth than other existing general unitary operations circuits already proposed in the literature \cite{Reck94,clements2016optimal,tabia2012experimental}. Other conclusion that arises from this result, is that the effect of the beam splitter that composes $\hat{T}_2$ solely exchanges the photon beam carried by the 2nd and 3rd paths, as seen in Eq.~\eqref{swap}. Except by this exchange, the BS's that acts non-trivially operates only between the 1st and 2nd, and the 3rd and 4th, separately. This structure will inspire the design of circuits for general qudit states.
		
	\section{Minimum quantum state tomography for qutrits}
	\label{sec:circuitsqutrit}
		
	For the qutrit, we approached the circuit's design in a different manner. The strategy to find a circuit that perform qubit's tomography starts with Eq.~\eqref{portaU}, which lead to the solution of a $4\times4$ matrix equation for finding the BS specifications. Applying the same procedure in the qutrit case would result in the solution of a $9 \times 9$ matrix equation. Alternatively, we faced this problem by analyzing the POVM operators and the desired outcomes for the qutrit tomography, being able to reduce the problem to the search of a $3 \times 3$ matrix equation solution. The qutrit tomography is realized via equidistant states \cite{Pimenta13,PaivaSanchez}.

	The SIC-POVM elements for the qutrit case are of the form
	\begin{equation}\label{purestates}
	    \hat{E}_{ml} = \frac{1}{3}| \psi_{ml}\rangle \langle \psi_{ml} | , 
	\end{equation}
	with $m,l=0,1,2$ and \cite{PaivaSanchez}
	\begin{subequations} 
		\begin{align}		
		|\psi_{00}\rangle & = \mu| 0 \rangle + \eta | 2 \rangle,\\
		|\psi_{01}\rangle & = \kappa| 0 \rangle + \kappa | 2 \rangle,  \\
		|\psi_{02}\rangle & = \eta| 0 \rangle + \mu | 2 \rangle,  \\	      
		|\psi_{10}\rangle & = \eta| 0 \rangle + \mu | 1 \rangle,   \\
		|\psi_{11}\rangle & = \kappa| 0 \rangle + \kappa | 1 \rangle, \\ 
		|\psi_{12}\rangle & = \mu| 0 \rangle + \eta | 1 \rangle,  \\
		|\psi_{20}\rangle & = \eta| 1 \rangle + \mu | 2 \rangle,  \\
		|\psi_{21}\rangle & = \kappa| 1 \rangle + \kappa | 2 \rangle,  \\
		|\psi_{22}\rangle & = \mu| 1 \rangle + \eta | 2 \rangle,
		\end{align}
	\end{subequations}
	 where $\kappa = 1/\sqrt{2}$, $\mu= e^{2\pi i / 3}/\sqrt{2}$, and $\eta = e^{- 2\pi i / 3}/\sqrt{2}$, such that $\kappa + \mu + \eta = 0$, ensuring that $\sum_{ml} \hat{E}_{ml} = \mathds{1}$. Keeping in mind Eq.~\eqref{barneteq}, we use the fact that the POVM elements are of the form of Eq.~\eqref{purestates} to define $a^{p}_{ml}$ as the expansion coefficients of $\ket{\psi_{ml}}$ as
	\begin{equation} \label{psiml}
	    \frac{1}{\sqrt{3}}\ket{\psi_{ml}} = a^{0}_{ml} \ket{0} + a_{ml}^1 \ket{1} + a_{ml}^2 \ket{2},
	\end{equation}
	where $\bra{p} \hat{E}_{ml} \ket{q} = (a_{ml}^p)^* a_{ml}^q$, with $p,q=0,1,2$. By substituing Eq.~\eqref{psiml} in Eq.~\eqref{purestates} and comparing $\hat{E}_{ml}$ with Eq.~\eqref{barneteq}, we arrive at an expression for the matrix elements of $\hat{U}$ that allow us to implement the desired POVM elements in the extended Hilbert space of dimension 9
	\begin{equation}
	\bra{ml} \hat{U} \ket{q0} = a_{ml}^q.
	\end{equation}
	
	\begin{figure}[]
		\centering
		\includegraphics[height=5.7cm]{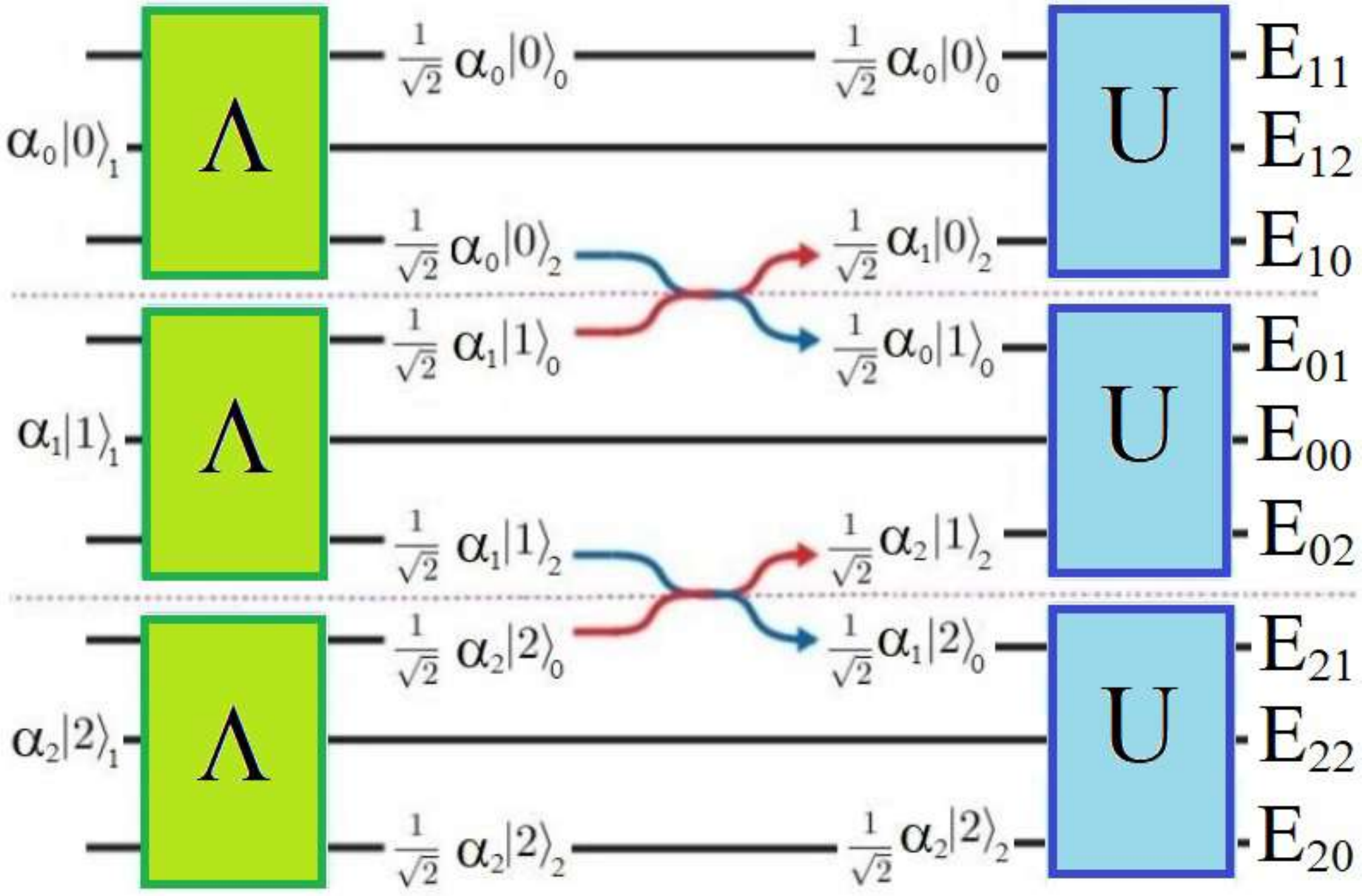}
		\caption{Schematic drawing showing the quantum operations of the proposed photonic circuit for quantum tomography of qutrits. The pink dotted lines separate the three sectors of the circuit, that can be analyzed separately. The waveguides are represented by the black lines and labeled by the vectors $\ket{i}_j$, where $i,j=0,1,2$. The state vectors $\alpha_i\ket{i}_1$ indicate the waveguides where a possible $(\alpha_0,\alpha_1,\alpha_2)^T$ input qutrit path state is coupled to the circuit. The $\hat{\Lambda}$ operation is responsible for diffusing the initial state to the first and third path in each sector. The red and blue arrows portray the permutations between the $3$rd and $4$th paths and between the $6$th and $7$th ones, respectively. This permutation, performed by $1:0$ BS, exchanges the coefficients of some base vectors. Finally, these paths are connected to the last circuit piece, described by a $3\times3$ $\hat{U}$ matrix, to resume the tomographic implementation. Analogous descriptions can be done for the others two sectors. A detector count in one of the nine circuit outputs is proportional to the implementation probability of a POVM element $\hat{E}_{ij}$ $(i,j=0,1,2)$.}
		\label{pedacoqutrit}
	\end{figure}	
	
	\begin{figure}[]
		\centering
		\includegraphics[height=4cm]{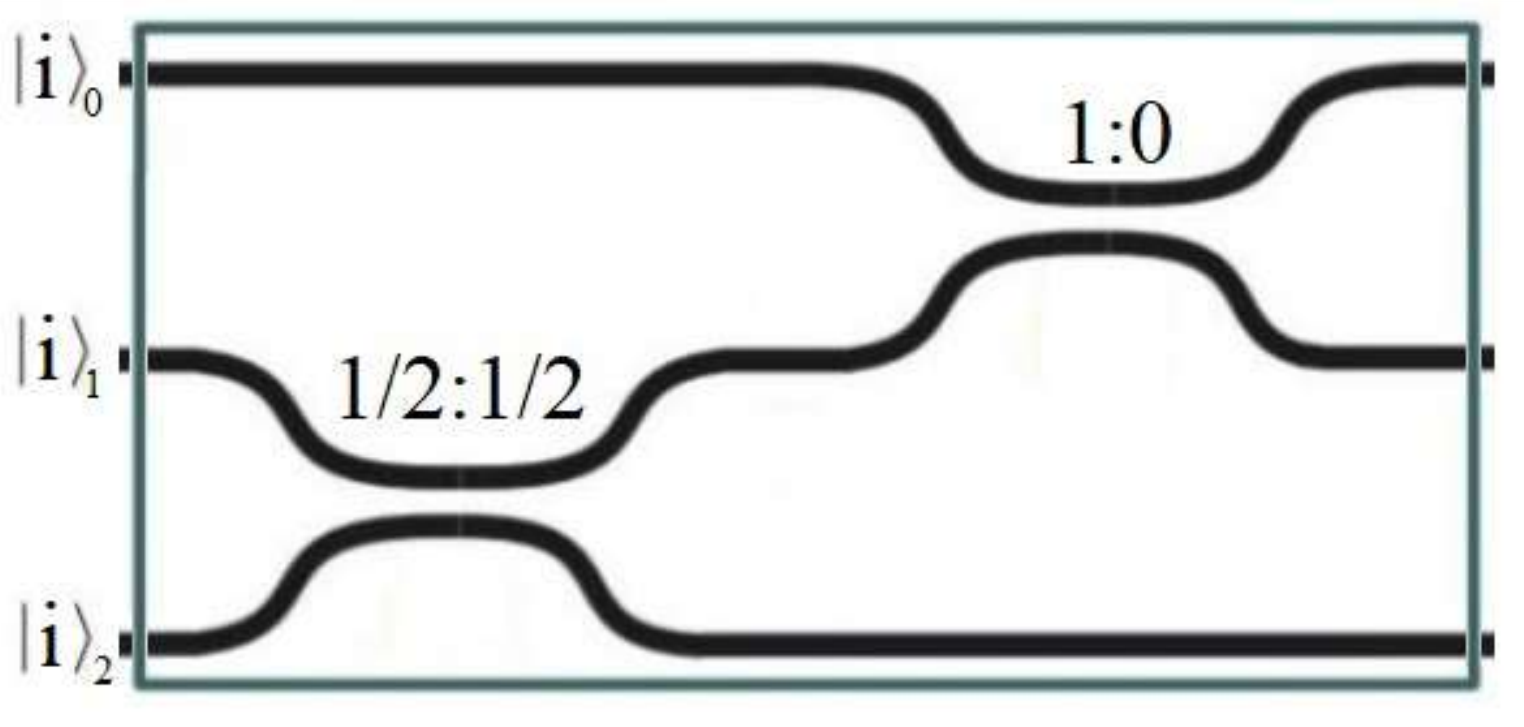}
		\caption{Circuit representation of the $\hat{\Lambda}$ operation for the implementation of the qutrit tomography. We have a $1/2:1/2$ BS  that splits the input mode in path 2 in two output modes in paths 2 and 3 (labelled as $\ket{i}_1$ and $\ket{i}_2$, respectively). A $1:0$ BS between paths 1 and 2 switches these paths.}
		\label{qutrit}
	\end{figure}
	
	Inspired by the solution for the tomography of qubits, we conceive an interferometer that is formed by three sectors, shown in Fig.~\ref{pedacoqutrit}, separated by the pink dotted line. Due to the symmetry in the POVM, it is convenient to perform the tomography measurements, such that, in the $k$th sector, the three outcomes result from the application of only $\hat{E}_{kl}$ $(l=0,1,2)$, already defined in Eq.~\eqref{purestates}. Our goal is that the photon count in the detector coupled to one of the three exits of the zeroth, first and second sector is proportional to the probability of implementing the POVM elements $\hat{E}_{0l}$, $\hat{E}_{1l}$ or $\hat{E}_{2l}$, respectively.
	
	The prepared photonic path state is such that the photon can enter in one of the nine input ports of the interferometer, as shown in Fig.~\ref{pedacoqutrit}. A photon in a qutrit path state represented by the $\ket{0}$, $\ket{1}$ and $\ket{2}$ base vectors can enter in the second or in the fifth or in the eighth input circuit ports. Since the photonic circuit has nine input ports, it is clear that the Hilbert space for the photonic path states increases from $N=3$ to $N=9$. We label the path states in the $N=9$ extended Hilbert space as $\ket{i}_j$, where $i,j=0,1,2$. The nine photonic paths, from top to bottom, are labelled as: $\ket{0}_0$,$\ket{0}_1$,$\ket{0}_2$, $\ket{1}_0$,$\ket{1}_1$,$\ket{1}_2$, $\ket{2}_0$,$\ket{2}_1$ and $\ket{2}_2$. Fig.~\ref{pedacoqutrit} shows the quantum operations schematically in the extended Hilbert space. 

	Let's suppose we have a normalized input state $\ket{\psi}=\sum_{i=0}^{2}\alpha_i\ket{i}_1$. After the first operation, represented by the green rectangles, the photon path state becomes $\ket{\psi'}=\sum_{i=0}^{2}\alpha_i\left(\ket{i}_0+\ket{i}_2\right)/\sqrt{2}$. The second transformation in Fig.~\ref{pedacoqutrit} switches some of the coefficients in $\ket{\psi'}$ to generate the state $\ket{\psi''}=\sum_{i=0}^{2}\left(\beta_i\ket{i}_0+\gamma_i\ket{i}_2\right)/\sqrt{2}$, where $\beta_0=\beta_1=\alpha_0$, $\beta_2=\gamma_0=\alpha_1$ and $\gamma_1=\gamma_2=\alpha_2$. The states $\ket{\psi}$ and $\ket{\psi''}$ can be represented by vectors such that the first and second state transformations in Fig.~\ref{pedacoqutrit} produces	
		
	\begin{equation}
		\left[
		\begin{array}{ccc}
		0 \\ \alpha_0 \\ 0 \\ 0 \\ \alpha_1 \\ 0 \\ 0 \\ \alpha_2 \\ 0 \\		  
		\end{array} 
		\right] \quad\longmapsto\quad \dfrac{1}{\sqrt{2}}
		\left[
		\begin{array}{ccc}
		\alpha_0 \\ 0 \\ \alpha_1 \\ \alpha_0 \\ 0 \\ \alpha_2 \\ \alpha_1 \\ 0 \\ \alpha_2 \\ 		  
		\end{array} 
		\right].
	\end{equation}
	
	\begin{figure}[]
	\centering
	\includegraphics[height=2.1cm]{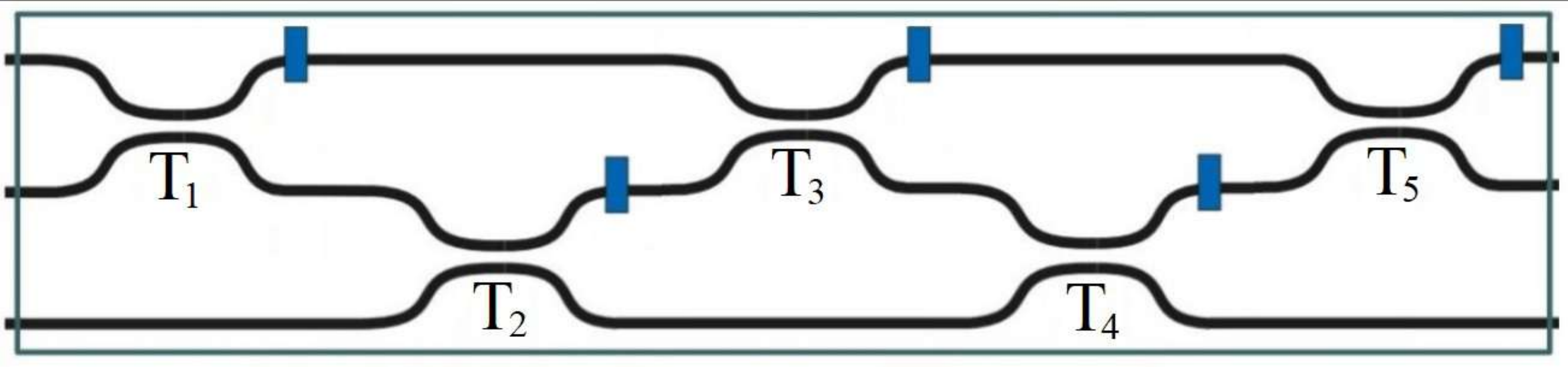}
	\caption{Circuit representation of the $\hat{U}$ matrix for the implementation of the qutrit tomography. The blue rectangles are phase shifters.}
	\label{figuraU}
\end{figure} 		
	
	Notice that the circuit design is such that, in the first, second, and third sectors, we create state superpositions of the original qutrit base states $\{\ket{0}$, $\ket{1}\}$, $\{\ket{0}$, $\ket{2}\}$ and $\{\ket{1}$, $\ket{2}\}$, respectively. The arrangement of beam splitters which implements what is outlined by the green rectangles in Fig.~\ref{pedacoqutrit} is shown in Fig.~\ref{qutrit}. The optical couplers (BS) that carry the quantum operation described in Fig.~\ref{qutrit} are $1/2:1/2$ BS ($t=0.5$ and $r=0.5$) and $1:0$ BS ($t=1$ and $r=0$). The second transformation shown in Fig.~\ref{pedacoqutrit}, the permutations between 3rd and 4th paths and 6th and 7th paths, is performed by $1:0$ BSs.
	
	The last part of Fig.~\ref{pedacoqutrit} shows the applications of unitary operations $\hat{U}$ in the first, second and third sectors. Notice that the $\hat{U}$ operations only apply to the vector states of their respective sectors. Therefore, the quantum operation described by a $9\times9$ matrix in the extended Hilbert space of 9 dimensions is composed by three $3\times3$ $\hat{U}$ matrices forming three blocks matrices with all the other elements of the $9\times9$ matrix equal to zero, i.e., 
	\begin{equation}
		\hat{O}=\left[
		\begin{array}{ccc}
		\hat{U} &  0 & 0 \\ 
		 0 & \hat{U} & 0 \\ 
		 0 & 0 & \hat{U}  \\ 	  
		\end{array} 
		\right].
	\end{equation}

The circuit scheme for $\hat{U}$ appears in detail in Fig.~\ref{figuraU}. Hence, we must establish the coefficients $r_j$, $t_j$, and the phases $\phi_j$. In order to achieve this, we are going to obtain the matrix from the inputs and outputs vectors available. The desired effect of the operations $\hat{E}_{0l}$, $\hat{E}_{1l}$ and $\hat{E}_{2l}$ in the first, second and third sectors of the circuit, respectively, leads us to
\begin{equation}
	\hat{U}\left[
	\begin{array}{ccc}
	\alpha_i/\sqrt{2} \\ 0 \\ \alpha_j/\sqrt{2} 
	\end{array} 
	\right]=\dfrac{1}{\sqrt{6}}
	\left[
	\begin{array}{ccc}
	\alpha_i+\alpha_j \\ \alpha_ie^{2i\pi/3}+\alpha_je^{-2i\pi/3} \\ \alpha_ie^{-2i\pi/3}+\alpha_je^{2i\pi/3}
	\end{array} 
	\right],
	\label{u}
\end{equation}
with $i=0,1$, $j=1,2$ and $i<j$. The unitary $\hat{U}$ that satisfies Eq.~\eqref{u} is
	\begin{equation}
	\hat{U} = \frac{1}{\sqrt{3}}
	\left[
	\begin{array}{ccc}
	1       & 1       & 1 \\ 
	e^{2i\pi/3}   & 1   & e^{-2i\pi/3} \\ 
	e^{-2i\pi/3}       & 1       & e^{2i\pi/3} 
	\end{array} 
	\right],
	\label{operana}
	\end{equation}
	that corresponds to the circuit in Fig.~\ref{figuraU}. We managed therefore to reduce the problem from a $9 \times 9$ matrix equation to a problem of finding coefficients for a $3 \times 3$ matrices. Performing this calculation (more details in Appendix B), we obtain the values of the coefficients and phases:	
	\begin{subequations}
		\begin{align}
		r_1 & = 1, & t_1 & = 0, & \phi_1 & = -2\pi/3, \\
		r_2 & = 0, & t_2 & = 1,  & \phi_2 & = -\pi/3, \\
		r_3 & = 1/2,  & t_3 & = 1/2, & \phi_3 & = -\pi/2, \\
		r_4 & = 1/3, & t_4 & = 2/3, & \phi_4 & = 0, \\
		r_5 & = 1/2, & t_5 & = 1/2, & \phi_5 & = \pi.
		\end{align}
	\end{subequations}

	The result presented in Eq. (26) shows that the first beam splitter of the scheme of Fig.~\ref{figuraU} can be discarded, since there is no transmittance. The phase shifter $\phi_5$ can also be discarded, since it is at the end of the circuit and does not interfere in the photon counting. We conclude that the interferometer that implements the quantum path state tomography in the one-qutrit state is formed by twenty beam splitters and nine phase shifters. As in the one-qubit case, our proposal for one-qutrit systems has less optical elements and a smaller optical depth than other existing proposals in the literature \cite{Reck94,clements2016optimal,tabia2012experimental}. 
	
	We also summarize in Table~\ref{table} the number of optical elements necessary for the implementation of each proposal of photonic circuits for the quantum tomography task. It is worth emphasizing that Reck's and Clements's proposals use Mach-Zehnder interferometers as building blocks of the circuits, while our proposal uses single beam splitters. 
	
	\begin{table}[]
		\caption{Number of optical elements required for the implementation of minimum quantum path state tomography in photonic circuits according to our proposal and others proposals already known in the literature\cite{Reck94,clements2016optimal,tabia2012experimental}.}
		\label{table}
		\begin{tabular}{|c|c|c|c|c|}
			\hline
			\multicolumn{1}{|l|}{} & \textbf{\begin{tabular}[c]{@{}c@{}}Reck's \\ proposal\end{tabular}} & \textbf{\begin{tabular}[c]{@{}c@{}}Clements's \\ proposal\end{tabular}} & \textbf{\begin{tabular}[c]{@{}c@{}}Tabia's \\ proposal\end{tabular}} & \textbf{\begin{tabular}[c]{@{}c@{}}Our \\ proposal\end{tabular}} \\ \hline
			\textbf{Qubit}         & 12                                                                  & 12                                                                      & 7                                                                    & 6                                                                \\ \hline
			\textbf{Qutrit}        & 72                                                                  & 72                                                                      & 44                                                                   & 29                                                               \\ \hline			
		\end{tabular}
	\end{table}

	\section{Generalization for $N$-dimensions}
	\label{sec:circuitsNd}
	
	In the previous section, we use the results of Paiva-S\'anchez \cite{PaivaSanchez} about equidistant states to build the SIC-POVM elements that implement a quantum state tomography of path photonic qutrits. One remarkable feature of this approach is that it is not limited to the dimension of the input state. The relation for equidistant states suitable for the tomography of an $N$-dimensional quantum system is \cite{PaivaSanchez}: 
	\begin{equation}
	\ket{\psi_{l}}=\dfrac{1}{\sqrt{N-1}}\sum_{\substack{k=0 \\ k\neq 1}}^{N-1}e^{2i\pi l\left(k-1\right)/N}\ket{k},
	\label{equidistante2}
	\end{equation}
	where $l=0,1,...,N-1$. From Eq.~\eqref{equidistante2}, we can to obtain $N$ operators of the $N^2$ required for the quantum state tomography experiment. The others operators are obtained from states generated by the application of the operator $\hat{X}$ in $\ket{\psi_l}$ shown in Eq.~\eqref{equidistante2}, defined as:
	\begin{equation}
	\hat{X}\ket{k}=\ket{k\oplus 1}.
	\label{x}
	\end{equation}
	This operator $\hat{X}$ acts on a state $\ket{k}$ performing an addition modulo $N$. Thus, all equidistant states used in the construction of the POVM elements can be organized in sets $B_m(\psi)$, with $m=0,1,...,N-1$, defined as 
	\begin{equation}
	B_m(\psi)=\left\{\ket{\psi_{ml}}=\hat{X}^m\ket{\psi_l}\right\}
	\end{equation} 
	and these POVM elements are obtained through the relation	
	\begin{equation}
\hat{E}_{ml} = \dfrac{1}{N}| \psi_{ml}\rangle \langle \psi_{ml}|.
\end{equation}	

		As an example, we will define the base states for quantum state tomography of ququarts, that is, qudits with $N=4$. They are:
	\begin{subequations}
	\begin{align}
		\ket{\psi_{00}}&=\dfrac{1}{\sqrt{3}}\left(\ket{0}+\ket{2}+\ket{3}\right); \\
		\ket{\psi_{01}}&=\dfrac{1}{\sqrt{3}}\left(-i\ket{0}+i\ket{2}-\ket{3}\right); \\
		\ket{\psi_{02}}&=\dfrac{1}{\sqrt{3}}\left(-\ket{0}-\ket{2}+\ket{3}\right); \\
		\ket{\psi_{03}}&=\dfrac{1}{\sqrt{3}}\left(i\ket{0}-i\ket{2}-\ket{3}\right).
	\end{align}	
	\end{subequations}

	The other states ($\ket{\psi_{1l}}$, $\ket{\psi_{2l}}$ and $\ket{\psi_{3l}}$, with $l=0,1,2,3$) come from the application of the operator $\hat{X}$ one, two or three times.  
	
	The determination of the circuit design for the general case is made in the same way as it was done in the Section IV, not applicable for qubit case. The $N^2$ paths are divided in the $N$ sectors with $N$ paths each. Here, the three-dimensional conception of circuits facilitates the generalization task \cite{chaboyer2015tunable,meany2015laser,poulios2014quantum}. The assumption of a 3D waveguide writing technique becomes necessary because we have to create path states superposition between states of different sectors. For example, it is necessary to create superposition between the states of the first and the last sector in a SIC-POVM tomographic proposal. In a planar circuit, the optical depth would increase, as long as these states are connected, by a quadratic function in $N$. In the 3D waveguide writing technique, these states will be connected by $N-1$ beam splitters, that is, the optical depth is now a linear function in $N$. 

	In the three-dimensional circuit scheme, the paths can be organized in $N$ parallel vertical sectors, with $N$ paths in the vertical direction each, forming a square array of waveguides. By choosing the paths for entrance of the photons belonging to the same line, i.e., path $i_0$ in each sector $j$, we have as initial state
	\begin{equation}
		\ket{\psi}_j^{(0)}=\sum_{j=0}^{N-1}\alpha_{j}\ket{i_0j},
		\label{initial}
	\end{equation}
	where $\ket{i_0j}$ means path state in line $i_0$ and sector $j$. Our proposal consists in conceiving the circuit as a sequence of three unitary operations. The first one operates separately in each vertical sector $j$ performing the transformation
	\begin{equation}
		\alpha_{j}\ket{i_0j} \quad\longmapsto\quad \dfrac{\alpha_{j}}{\sqrt{N-1}}\sum_{i=0}^{N-2}\ket{ij},
	\end{equation}
	i.e., each component of the initial state $\alpha_j\ket{i_0j}$ is decomposed into $N-1$ components $\alpha_j\ket{ij}$, that will occupy $N-1$ paths of the vertical sector $j$, leaving, also by choice, the last path empty.
	
	The second operation realized by the circuit is a permutation betweens vertical sectors. Except the last layer, that did not receive one of the $N-1$ components at the end of the first operation, in all other layers the operation performed is an addiction modulo $N$ as follows
	\begin{equation}
		\hat{\varGamma}\ket{ij}=\ket{i}\ket{j\oplus i}, 
	\end{equation}
	where $i=0,1,...,N-2$ is the paths label. At the end of this second operation, we will have in each vertical sector $j$ the state
	\begin{equation}
		\ket{\psi}_j^{(2)}=\dfrac{1}{\sqrt{N-1}}\sum_{i=0}^{N-2}\alpha_{j-i}\ket{ij},
		\label{psi2}
	\end{equation}
 	where the operations with subindexes are carried out modulo $N$. The third operation that our proposed circuit performs is analogous to Eq.~\eqref{operana}. Its transforms the state $\ket{\psi}_j^{(2)}$ (Eq.~\eqref{psi2}) in a final state where the square module of its expansion coefficients are igual to the probabilities of implement all POVM elements in the qudit Hilbert space. By possession of these probabilities, we reconstruct the initial state density matrix \cite{PaivaSanchez}.
	
	If the possibility of usage of a 3-dimensional array is at hand, our circuit becomes considerably more efficient that one derived directly from \cite{Reck94} or \cite{clements2016optimal}. As the minimal state tomography uses at least $N^2$ outputs, a general circuit would have a complexity of $O(N^4)$ beam splitters. In our realization, the procedure is divide into three parts, each acting in parallel over at maximum $N$ paths. This guarantees a complexity of $O(N^3)$. Moreover, each part may simplify separately, since they act differently, allowing an increase of efficiency through the detalied analysis of each part. The modular property of the circuit is the main feature of our proposed implementation of the circuit for minimal state tomography. 
	
	There is a particularity in the even-dimension case, reported in Ref. \cite{PaivaSanchez}. It is shown that the method exposed in this paper cannot determine the imaginary part of some elements of the system's density matrix $\hat{\rho}$. For that, one more step is needed, the application of the following operator:
	\begin{equation}
	\hat{\Xi}=\sum_{k=0}^{N/2-1}\ket{k}\bra{k}+i\sum_{N/2}^{N-1}\ket{k}\bra{k},
	\label{ureal}
	\end{equation}
	which generates a new density matrix $\hat{\rho}'=\hat{\Xi}\hat{\rho}\hat{\Xi}^\dag$. The advantage about this strategy is that the imaginary parts, impossible to be determined, are exchanged with the real parts of the same matrix elements, being now possible to be obtained. Thus, to reconstruct the initial density matrix of even-dimensional systems, it is necessary to realize the quantum state tomography in $\hat{\rho}$ and $\hat{\rho}'$, increasing from $N^2$ to $3N^2/2$ the number of measurements. 
		
	\section{Conclusion}
	\label{sec:conclusao}
	
	In this paper, we were able to determine the design of integrated photonic circuits that perform quantum path state tomography for qubits and qutrits. In these schemes, a POVM is implemented by doing an application of Naimark's theorem by extending the qudit Hilbet space, performing unitary transformations and projective measurements in this extended space. The parameters of the optical interferometer elements (beam splitters and phase shifters) were accurately calculated by using methods already described in the literature. Only three different types of beam splitters appeared on the proposed circuits, which may facilitate their manufacture. Our designs requires less optical components than alternatives general unitary transformations circuits and a smaller optical depth when compared with the optical depth of the circuits discussed in \cite{Reck94,clements2016optimal}. A general recipe also was presented for determining the circuit to be used for dimensions higher than three. In all cases, our interferometers are more compact than those already exposed in other works.

	
	\begin{acknowledgements} This research was supported by the Brazilian agencies CNPq - Conselho Nacional de Desenvolvimento Cient\'{\i}fico e Tecnol\'ogico, Capes - Coordena\c{c}\~ao de Aperfei\c{c}oamento de Pessoal de N\'{\i}vel Superior, Fapemig - Funda\c{c}\~ao de Amparo \`a Pesquisa do Estado de Minas Gerais and INCT-IQ - Instituto Nacional de Ci\^encia e Tecnologia de Informa\c{c}\~ao Qu\^antica.
	\end{acknowledgements}

	
	\appendix
	
	\section{Calculations for the Qubit Photonic Circuit}
	
	In Sec. II are defined the POVM elements that realize a full quantum state tomography in one-qubit systems, namely
	\begin{equation}\label{POVM1}
	\hat{E}_{1} = \frac{1}{2}
	\left[
	\begin{array}{cc}
	2/3         & -i\sqrt{2}/3 \\ 
	i\sqrt{2}/3 & 1/3
	\end{array}
	\right], 
	\end{equation}
	\begin{equation}\label{POVM2}
	\hat{E}_{2} = \frac{1}{2}
	\left[
	\begin{array}{cc}
	2/3           & i\sqrt{2}/3 \\
	-i\sqrt{2}/3  & 1/3
	\end{array}
	\right],  
	\end{equation}
	\begin{equation}\label{POVM3} 
	\hat{E}_{3} = \frac{1}{2}
	\left[
	\begin{array}{cc}
	1/3         & -\sqrt{2}/3 \\
	-\sqrt{2}/3 & 2/3
	\end{array}
	\right], 
	\end{equation}
	\begin{equation}\label{POVM4}
	\hat{E}_{4} = \frac{1}{2}
	\left[
	\begin{array}{cc}
	1/3         & \sqrt{2}/3 \\ 
	\sqrt{2}/3  & 2/3
	\end{array}
	\right]. 
	\end{equation}
	
	On the other hand, the POVM elements written in terms of the parameters of the photonic circuit that we aim to determine are expressed in Eq.~\eqref{traco2}, where $\ket{m}$ and $\ket{l}$ are the state vectors of the subspace basis, principal system and \textit{ancilla}, respectively. Instead of labelling the outcomes in terms of the binary base $\ket{ml}$, we choose to label the output states in terms of the path number of the circuit, where we write $\hat{\rho}_k=\ket{k}\bra{k}$. Therefore, we make the identification:
	\begin{equation}
	\begin{split}
	m=0, l=0 \quad\rightarrow\quad k=1, \\
	m=0, l=1 \quad\rightarrow\quad k=2, \\
	m=1, l=0 \quad\rightarrow\quad k=3, \\
	m=1, l=1 \quad\rightarrow\quad k=4.
	\end{split}
	\end{equation}
	
	Moreover, these operators are simplified if we define the beam splitter in $\hat{T}_2$ as a $1:0$ BS, so that $r_3=0$ and $t_3=1$. With this, the operators are reduced to:	
	\begin{equation} \label{opera12}
	\hat{E}_1=\left[
	\begin{array}{cccc}
	r_4t_1 & e^{-i\theta}\sqrt{r_2r_4t_1t_4} \\
	e^{i\theta}\sqrt{r_2r_4t_1t_4} & r_2t_4   
	\end{array}\right],
	\end{equation}
	\begin{equation} \label{opera22}
	\hat{E}_2=\left[
	\begin{array}{cccc}
	t_1t_4 & -e^{-i\theta}\sqrt{r_2r_4t_1t_4} \\
	-e^{i\theta}\sqrt{r_2r_4t_1t_4} & r_2r_4   
	\end{array}\right],
	\end{equation}
	\begin{equation} \label{opera32}
	\hat{E}_3=\left[
	\begin{array}{cccc}
	r_1r_5 & -\sqrt{r_1r_5t_2t_5} \\
	-\sqrt{r_1r_5t_2t_5} & t_2t_5
	\end{array}\right], \quad
	\end{equation}
	\begin{equation} \label{opera42}
	\hat{E}_4=\left[
	\begin{array}{cccc}
	r_1t_5 & \sqrt{r_1r_5t_2t_5} \\
	\sqrt{r_1r_5t_2t_5} & r_5t_2
	\end{array}\right], \quad
	\end{equation}
	where $\theta=\phi_1-\phi_2-\phi_3$. The absence of the phases $\phi_4$ and $\phi_5$ in the operators $\hat{E}_k$ allows us to choose:
	\begin{equation}
	\phi_4 = 0,\qquad\qquad  \phi_5=0.
	\end{equation}	
	
	When comparing $\hat{E}_1$ (Eq.~\eqref{POVM1}) with $\hat{E}_1$ (Eq.~\eqref{opera12}) and $\hat{E}_2$ (Eq.~\eqref{POVM2}) with $\hat{E}_2$ (Eq.~\eqref{opera22}), it is concluded that:
	\begin{equation} \label{r4t4}
	r_4=t_4 \quad \Rightarrow \quad r_4=1/2, \quad t_4=1/2.
	\end{equation}
	
	Similarly, when comparing $\hat{E}_3$ (Eq.~\eqref{POVM3}) with $\hat{E}_3$ (Eq.~\eqref{opera32}) and $\hat{E}_4$ (Eq.~\eqref{POVM4}) with $\hat{E}_4$ (Eq.~\eqref{opera42}), it is concluded that:
	\begin{equation}
	r_5=t_5 \quad \Rightarrow \quad r_5=1/2, \quad t_5=1/2.
	\end{equation}
	
	Therefore, the operators acquire simpler forms:	
	\begin{equation} \label{opera13}
	\hat{E}_1=\dfrac{1}{2}\left[
	\begin{array}{cccc}
	t_1 & e^{-i\theta}\sqrt{r_2t_1} \\
	e^{i\theta}\sqrt{r_2t_1} & r_2   
	\end{array}\right],
	\end{equation}
	\begin{equation} \label{opera23}
	\hat{E}_2=\dfrac{1}{2}\left[
	\begin{array}{cccc}
	t_1 & -e^{-i\theta}\sqrt{r_2t_1} \\
	-e^{i\theta}\sqrt{r_2t_1} & r_2   
	\end{array}\right],
	\end{equation}
	\begin{equation} \label{opera33}
	\hat{E}_3=\dfrac{1}{2}\left[
	\begin{array}{cccc}
	r_1 & -\sqrt{r_1t_2} \\
	-\sqrt{r_1t_2} & t_2
	\end{array}\right],
	\end{equation}
	\begin{equation} \label{opera43}
	\hat{E}_4=\dfrac{1}{2}\left[
	\begin{array}{cccc}
	r_1 & \sqrt{r_1t_2} \\
	\sqrt{r_1t_2} & t_2
	\end{array}\right].
	\end{equation}
	
	Make the same comparisons explained above, we obtain the last parameters of the circuit:
	\begin{equation}
	\begin{array}{rcl}
	r_1 = 1/3,\qquad & t_1 = 2/3,\qquad  \\
	r_2 = 1/3,\qquad & t_2 = 2/3,\qquad   \\
	\end{array}
	\theta=\pi/2.
	\end{equation}
	
	\section{Calculations for the $\hat{U}$ Matrix of the Qutrit Photonic Circuit}
	
	We are interested to find the parameters necessary for carrying the operator $\hat{U}$ shown in Eq.~\eqref{operana}. The circuit representation of the $\hat{U}$ matrix is shown in Fig.~\ref{figuraU}. The operations that make up this circuit are:
		
	\begin{equation}
	\hat{T}_a =
	\left[
	\begin{array}{cccc}
	e^{i \phi_{a}} \sqrt{r_{a}} & e^{i \phi_{a}} \sqrt{t_{a}}      & 0  \\
	\sqrt{t_{a}}                & -\sqrt{r_{a}}   & 0  \\
	0 & 0            & 1          \\
	\end{array}
	\right], 
	\end{equation}
	\begin{equation}
	\hat{T}_b =
	\left[
	\begin{array}{cccc}
	1 & 0                           & 0                            \\
	0 & e^{i \phi_{b}} \sqrt{r_{b}} & e^{i \phi_{b}} \sqrt{t_{b}}  \\
	0 & \sqrt{t_{b}}                & -\sqrt{r_{b}}               \\
	\end{array}
	\right],
	\end{equation}
	where $a=1,3,5$ and $b=2,4$. The entire circuit is represented by the matrix:	
	\begin{equation}\label{circuitoqutrit}
	\hat{U}_{qutrit}=\hat{T}_5 \cdot \hat{T}_{4} \cdot \hat{T}_3 \cdot \hat{T}_{2} \cdot \hat{T}_1.
	\end{equation}	

	Analogous to the qubit case, it is possible to simplify the matrix $\hat{U}_{qutrit}$ by defining $r_1=1$, $t_1=0$, $r_2=0$ and $t_2=1$. This reduces the matrix elements to the following quantities	
	
	\begin{equation} \nonumber
	\bra{0}\hat{U}_{qutrit}\ket{0}=e^{i(\phi_1+\phi_5)}(e^{i\phi_3}\sqrt{r_3r_5}+e^{i\phi_4}\sqrt{r_4t_3t_5}), 
	\end{equation}	
	\begin{equation}\nonumber
	\bra{0}\hat{U}_{qutrit}\ket{1}=-e^{i(\phi_4+\phi_5)}\sqrt{t_4t_5},
	\end{equation}	
	\begin{equation}\nonumber
	\bra{0}\hat{U}_{qutrit}\ket{2}=e^{i(\phi_2+\phi_5)}(e^{i\phi_3}\sqrt{r_5t_3}-e^{i\phi_4}\sqrt{r_3r_4t_5}), 
	\end{equation}
	\begin{equation}\nonumber
	\bra{1}\hat{U}_{qutrit}\ket{0}=e^{i\phi_1}\left(-e^{i\phi_4}\sqrt{r_4r_5t_3}+e^{i\phi_3}\sqrt{r_3t_5}\right), 
	\end{equation}
	\begin{equation}\nonumber
	\bra{1}\hat{U}_{qutrit}\ket{1}=-e^{i\phi_4}\sqrt{r_5t_4}, 
	\end{equation}
	\begin{equation}\nonumber
	\bra{1}\hat{U}_{qutrit}\ket{2}=e^{i\phi_2}\left(e^{i\phi_4}\sqrt{r_3r_4r_5}+e^{i\phi_3}\sqrt{t_3t_5}\right), 
	\end{equation}
	\begin{equation}\nonumber
	\bra{2}\hat{U}_{qutrit}\ket{0}=e^{i\phi_1}\sqrt{t_3t_4}, 
	\end{equation}
	\begin{equation}\nonumber
	\bra{2}\hat{U}_{qutrit}\ket{1}=\sqrt{r_4}, 
	\end{equation}
	\begin{equation}\nonumber
	\bra{2}\hat{U}_{qutrit}\ket{2}=-e^{i\phi_2}\sqrt{r_3t_4}. 
	\end{equation}
	
	A first comparison between the elements of the matrix $\hat{U}_{qutrit}$ and those of the matrix $\hat{U}$ in Eq.~\eqref{operana} allows us to determine the following parameters	
	\begin{equation}
	\begin{array}{rcl}
	r_3 = 1/2,\qquad & t_3 = 1/2,\qquad & \phi_1 = -2\pi/3, \\
	r_4 = 1/3,\qquad & t_4 = 2/3,\qquad & \phi_2 = -\pi/3.  
	\end{array}
	\label{param3}
	\end{equation}	
	
	With the parameters shown in Eq.~\eqref{param3}, the matrix elements $\bra{2}\hat{U}_{qutrit}\ket{j}$ ($j=0,1,2$), are determined completely. The parameters $\phi_j$ $(j=3,4,5)$, $r_5$ and $t_5$ of the remaining elements are still not determined, as shown below	
	\begin{equation}\nonumber
	\bra{0}\hat{U}_{qutrit}\ket{0}=e^{i(\phi_5-2\pi/3)}(e^{i\phi_3}\sqrt{r_5/2}+e^{i\phi_4}\sqrt{t_5/6}), 
	\end{equation}		
	\begin{equation}\nonumber
	\bra{0}\hat{U}_{qutrit}\ket{1}=-e^{i(\phi_4+\phi_5)}\sqrt{2t_5/3},
	\end{equation}
	\begin{equation}\nonumber
	\bra{0}\hat{U}_{qutrit}\ket{2}=e^{i(\phi_5-\pi/3)}(e^{i\phi_3}\sqrt{r_5/2}-e^{i\phi_4}\sqrt{t_5/6}), 
	\end{equation}		
	\begin{equation}\nonumber
	\bra{1}\hat{U}_{qutrit}\ket{0}=e^{-2i\pi/3}(-e^{i\phi_4}\sqrt{r_5/6}+e^{i\phi_3}\sqrt{t_5/2}), 
	\end{equation}	
	\begin{equation}\nonumber
	\bra{1}\hat{U}_{qutrit}\ket{1}=-e^{i\phi_4}\sqrt{2r_5/3}, 
	\end{equation}	
	\begin{equation}\nonumber
	\bra{1}\hat{U}_{qutrit}\ket{2}=e^{-i\pi/3}(e^{i\phi_4}\sqrt{r_5/6}+e^{i\phi_3}\sqrt{t_5/2}).
	\end{equation}	
			
	When comparing again the elements $\bra{i}\hat{U}_{qutrit}\ket{j}$ ($i=0,1$, $j=0,1,2$) and the respective elements of the $\hat{U}$ matrix in Eq.~\eqref{operana}, we are able to determine the remaining parameters	
	\begin{equation}
	\begin{split}
	r_5 = 1/2,\qquad  t_5 = 1/2, \qquad\qquad \\
	\phi_3 = 1/3,\qquad  \phi_4 = 2/3,\qquad \phi_5 = -\pi/3.  
	\end{split}
	\end{equation}
		
	Thus, we obtained $\hat{U}_{qutrit}=\hat{U}$ as we wanted.
	
    
\bibliographystyle{unsrt}
\bibliography{Referencias}

\begin{thebibliography}{10}

\bibitem{rungta2001qudit}
P~Rungta, WJ~Munro, K~Nemoto, P~Deuar, Gerard~J Milburn, and CM~Caves.
\newblock Qudit entanglement.
\newblock In {\em Directions in Quantum Optics}, pages 149--164. Springer,
  2001.

\bibitem{thew2004bell}
Robert~Thomas Thew, Antonio Acin, Hugo Zbinden, and Nicolas Gisin.
\newblock Bell-type test of energy-time entangled qutrits.
\newblock {\em Physical review letters}, 93(1):010503, 2004.

\bibitem{de2002creating}
Hugues De~Riedmatten, Ivan Marcikic, Hugo Zbinden, and Nicolas Gisin.
\newblock Creating high dimensional time-bin entanglement using mode-locked
  lasers.
\newblock {\em Quant. Inf. Comput. 2}, pages 425--433, 2002.

\bibitem{thew2002qudit}
RT~Thew, Kae Nemoto, Andrew~G White, and William~J Munro.
\newblock Qudit quantum-state tomography.
\newblock {\em Physical Review A}, 66(1):012303, 2002.

\bibitem{neves2005generation}
Leonardo Neves, G~Lima, JG~Aguirre G{\'o}mez, CH~Monken, C~Saavedra, and
  S~P{\'a}dua.
\newblock Generation of entangled states of qudits using twin photons.
\newblock {\em Physical review letters}, 94(10):100501, 2005.

\bibitem{moreva2006realization}
EV~Moreva, GA~Maslennikov, SS~Straupe, and SP~Kulik.
\newblock Realization of four-level qudits using biphotons.
\newblock {\em Physical review letters}, 97(2):023602, 2006.

\bibitem{li2008generation}
Yongmin Li, Kuanshou Zhang, and Kunchi Peng.
\newblock Generation of qudits and entangled qudits.
\newblock {\em Physical Review A}, 77(1):015802, 2008.

\bibitem{Ursin17}
Fabian Steinlechner, Sebastian Ecker, Matthias Fink, Bo~Liu, Jessica Bavaresco,
  Marcus Huber, Thomas Scheidl, and Rupert Ursin.
\newblock {Distribution of high-dimensional entanglement via an intra-city
  free-space link}.
\newblock {\em Nature Communications}, 8:15971, jul 2017.

\bibitem{Torres12}
Martin Hendrych, Rodrigo Gallego, Michal Mi{\v{c}}uda, Nicolas Brunner, Antonio
  Ac{\'{i}}n, and Juan~P Torres.
\newblock {Experimental estimation of the dimension of classical and quantum
  systems}.
\newblock {\em Nature Physics}, 8:588, jun 2012.

\bibitem{kues2017chip}
Michael Kues, Christian Reimer, Piotr Roztocki, Luis~Romero Cort{\'e}s,
  Stefania Sciara, Benjamin Wetzel, Yanbing Zhang, Alfonso Cino, Sai~T Chu,
  Brent~E Little, et~al.
\newblock On-chip generation of high-dimensional entangled quantum states and
  their coherent control.
\newblock {\em Nature}, 546(7660):622, 2017.

\bibitem{wang2018multidimensional}
Jianwei Wang, Stefano Paesani, Yunhong Ding, Raffaele Santagati, Paul
  Skrzypczyk, Alexia Salavrakos, Jordi Tura, Remigiusz Augusiak, Laura
  Man{\v{c}}inska, Davide Bacco, et~al.
\newblock Multidimensional quantum entanglement with large-scale integrated
  optics.
\newblock {\em Science}, 360(6386):285--291, 2018.

\bibitem{walborn2006quantum}
SP~Walborn, DS~Lemelle, MP~Almeida, and PH~Souto Ribeiro.
\newblock Quantum key distribution with higher-order alphabets using spatially
  encoded qudits.
\newblock {\em Physical review letters}, 96(9):090501, 2006.

\bibitem{wang2017qudit}
Dong-Sheng Wang, David~T Stephen, and Robert Raussendorf.
\newblock Qudit quantum computation on matrix product states with global
  symmetry.
\newblock {\em Physical Review A}, 95(3):032312, 2017.

\bibitem{wang2018proof}
Shuang Wang, Zhen-Qiang Yin, HF~Chau, Wei Chen, Chao Wang, Guang-Can Guo, and
  Zheng-Fu Han.
\newblock Proof-of-principle experimental realization of a qubit-like
  qudit-based quantum key distribution scheme.
\newblock {\em Quantum Science and Technology}, 3(2):025006, 2018.

\bibitem{etcheverry2013quantum}
Sebastian Etcheverry, Gustavo Ca{\~n}as, ES~G{\'o}mez, WAT Nogueira,
  C~Saavedra, GB~Xavier, and Gustavo Lima.
\newblock Quantum key distribution session with 16-dimensional photonic states.
\newblock {\em Scientific reports}, 3:2316, 2013.

\bibitem{gedik2015computational}
Zafer Gedik, Isabela~Almeida Silva, Bar{\i}{\c{s}} {\c{C}}akmak, G{\"o}ktug
  Karpat, Edson Luiz~G{\'e}a Vidoto, Diogo de~Oliveira Soares-Pinto,
  FF~Fanchini, et~al.
\newblock Computational speed-up with a single qudit.
\newblock {\em Scientific reports}, 5:14671, 2015.

\bibitem{niu2018qudit}
Murphy~Yuezhen Niu, Isaac~L Chuang, and Jeffrey~H Shapiro.
\newblock Qudit-basis universal quantum computation using $\chi$ (2)
  interactions.
\newblock {\em Physical review letters}, 120(16):160502, 2018.

\bibitem{chau2015quantum}
HF~Chau.
\newblock Quantum key distribution using qudits that each encode one bit of raw
  key.
\newblock {\em Physical Review A}, 92(6):062324, 2015.

\bibitem{islam2017provably}
Nurul~T Islam, Charles Ci~Wen Lim, Clinton Cahall, Jungsang Kim, and Daniel~J
  Gauthier.
\newblock Provably secure and high-rate quantum key distribution with time-bin
  qudits.
\newblock {\em Science advances}, 3(11):e1701491, 2017.

\bibitem{pal2018entanglement}
Rajarshi Pal and Somshubhro Bandyopadhyay.
\newblock Entanglement sharing via qudit channels: Nonmaximally entangled
  states may be necessary for one-shot optimal singlet fraction and negativity.
\newblock {\em Physical Review A}, 97(3):032322, 2018.

\bibitem{agnew2011tomography}
Megan Agnew, Jonathan Leach, Melanie McLaren, F~Stef Roux, and Robert~W Boyd.
\newblock Tomography of the quantum state of photons entangled in high
  dimensions.
\newblock {\em Physical Review A}, 84(6):062101, 2011.

\bibitem{Pimenta13}
W.M. Pimenta, B.~Marques, T.O. Maciel, R.O. Vianna, A.~Delgado, C.~Saavedra,
  and S.~P{\'a}dua.
\newblock Minimum tomography of two entangled qutrits using local measurements
  of one-qutrit symmetric informationally complete positive operator-valued
  measure.
\newblock {\em Physical Review A}, 88(1):012112, 2013.

\bibitem{bent2015experimental}
N~Bent, H~Qassim, AA~Tahir, D~Sych, G~Leuchs, LL~S{\'a}nchez-Soto, E~Karimi,
  and RW~Boyd.
\newblock Experimental realization of quantum tomography of photonic qudits via
  symmetric informationally complete positive operator-valued measures.
\newblock {\em Physical Review X}, 5(4):041006, 2015.

\bibitem{ikuta2017implementation}
Takuya Ikuta and Hiroki Takesue.
\newblock Implementation of quantum state tomography for time-bin qudits.
\newblock {\em New Journal of Physics}, 19(1):013039, 2017.

\bibitem{martinez2019experimental}
D~Mart{\'\i}nez, MA~Sol{\'\i}s-Prosser, G~Ca{\~n}as, O~Jim{\'e}nez, A~Delgado,
  and G~Lima.
\newblock Experimental quantum tomography assisted by multiply symmetric states
  in higher dimensions.
\newblock {\em Physical Review A}, 99(1):012336, 2019.

\bibitem{gutierrez2012experimental}
AJ~Guti{\'e}rrez-Esparza, WM~Pimenta, B~Marques, AA~Matoso, S~P{\'a}dua, et~al.
\newblock Experimental characterization of two spatial qutrits using
  entanglement witnesses.
\newblock {\em Optics express}, 20(24):26351--26362, 2012.

\bibitem{gutierrez2014detection}
AJ~Guti{\'e}rrez-Esparza, WM~Pimenta, B~Marques, AA~Matoso, J~Sperling,
  W~Vogel, and S~P{\'a}dua.
\newblock Detection of nonlocal superpositions.
\newblock {\em Physical Review A}, 90(3):032328, 2014.

\bibitem{marques2015experimental}
B~Marques, AA~Matoso, WM~Pimenta, AJ~Guti{\'e}rrez-Esparza, MF~Santos, and
  S~P{\'a}dua.
\newblock Experimental simulation of decoherence in photonics qudits.
\newblock {\em Scientific reports}, 5:16049, 2015.

\bibitem{varga2018characterizing}
Juan Jos{\'e}~Miguel Varga, Lorena Reb{\'o}n, Q~Pears Stefano, and Claudio
  Iemmi.
\newblock Characterizing d-dimensional quantum channels by means of quantum
  process tomography.
\newblock {\em Optics letters}, 43(18):4398--4401, 2018.

\bibitem{Politi08}
Alberto Politi, Martin~J. Cryan, John~G. Rarity, Siyuan Yu, and Jeremy~L.
  O{\textquoteright}Brien.
\newblock Silica-on-silicon waveguide quantum circuits.
\newblock {\em Science}, 320(5876):646--649, 2008.

\bibitem{Marshall09}
Graham~D. Marshall, Alberto Politi, Jonathan C.~F. Matthews, Peter Dekker,
  Martin Ams, Michael~J. Withford, and Jeremy~L. O'Brien.
\newblock Laser written waveguide photonic quantum circuits.
\newblock {\em Opt. Express}, 17(15):12546--12554, Jul 2009.

\bibitem{shadbolt2012generating}
Peter~J Shadbolt, Maria~R Verde, Alberto Peruzzo, Alberto Politi, Anthony
  Laing, Mirko Lobino, Jonathan~CF Matthews, Mark~G Thompson, and Jeremy~L
  O'Brien.
\newblock Generating, manipulating and measuring entanglement and mixture with
  a reconfigurable photonic circuit.
\newblock {\em Nature Photonics}, 6(1):45, 2012.

\bibitem{o2009photonic}
Jeremy~L O'brien, Akira Furusawa, and Jelena Vu{\v{c}}kovi{\'c}.
\newblock Photonic quantum technologies.
\newblock {\em Nature Photonics}, 3(12):687, 2009.

\bibitem{caspani2017integrated}
Lucia Caspani, Chunle Xiong, Benjamin~J Eggleton, Daniele Bajoni, Marco
  Liscidini, Matteo Galli, Roberto Morandotti, and David~J Moss.
\newblock Integrated sources of photon quantum states based on nonlinear
  optics.
\newblock {\em Light: Science \& Applications}, 6(11):e17100, 2017.

\bibitem{smith2009phase}
Brian~J Smith, Dmytro Kundys, Nicholas Thomas-Peter, PGR Smith, and
  IA~Walmsley.
\newblock Phase-controlled integrated photonic quantum circuits.
\newblock {\em Optics Express}, 17(16):13516--13525, 2009.

\bibitem{Davis96}
K.~M. Davis, K.~Miura, N.~Sugimoto, and K.~Hirao.
\newblock Writing waveguides in glass with a femtosecond laser.
\newblock {\em Opt. Lett.}, 21(21):1729--1731, Nov 1996.

\bibitem{Streltsov01}
Alexander~M. Streltsov and Nicholas~F. Borrelli.
\newblock Fabrication and analysis of a directional coupler written in glass by
  nanojoule femtosecond laser pulses.
\newblock {\em Opt. Lett.}, 26(1):42--43, Jan 2001.

\bibitem{Walther12}
Al{\'{a}}n Aspuru-Guzik and Philip Walther.
\newblock {Photonic quantum simulators}.
\newblock {\em Nature Physics}, 8:285, apr 2012.

\bibitem{Koch12}
Andrew~A Houck, Hakan~E T{\"{u}}reci, and Jens Koch.
\newblock {On-chip quantum simulation with superconducting circuits}.
\newblock {\em Nature Physics}, 8:292, apr 2012.

\bibitem{harris2017quantum}
Nicholas~C Harris, Gregory~R Steinbrecher, Mihika Prabhu, Yoav Lahini, Jacob
  Mower, Darius Bunandar, Changchen Chen, Franco~NC Wong, Tom Baehr-Jones,
  Michael Hochberg, et~al.
\newblock Quantum transport simulations in a programmable nanophotonic
  processor.
\newblock {\em Nature Photonics}, 11(7):447, 2017.

\bibitem{kerman2017multiloop}
Andrew~J Kerman.
\newblock Multiloop interferometers for quantum information processing, May~18
  2017.
\newblock US Patent App. 15/354,275.

\bibitem{Fan08}
Fan Qiu-Bo.
\newblock Remote preparation of photon polarization states within a network.
\newblock {\em Chinese Physics Letters}, 25(1):20, 2008.

\bibitem{Yin11}
Wang Zhang-Yin.
\newblock Controlled remote preparation of a two-qubit state via an asymmetric
  quantum channel.
\newblock {\em Communications in Theoretical Physics}, 55(2):244, 2011.

\bibitem{wang2015controlled}
Chun Wang, Zhi Zeng, and Xi-Han Li.
\newblock Controlled remote state preparation via partially entangled quantum
  channel.
\newblock {\em Quantum Information Processing}, 14(3):1077--1089, 2015.

\bibitem{cao2016flexible}
Thi~Bich Cao, Ba~An Nguyen, et~al.
\newblock Flexible controlled joint remote preparation of an arbitrary
  two-qubit state via non-maximally entangled quantum channels.
\newblock {\em Advances in Natural Sciences: Nanoscience and Nanotechnology},
  7(2):025007, 2016.

\bibitem{Schaeff15}
Christoph Schaeff, Robert Polster, Marcus Huber, Sven Ramelow, and Anton
  Zeilinger.
\newblock Experimental access to higher-dimensional entangled quantum systems
  using integrated optics.
\newblock {\em Optica}, 2(6):523--529, 2015.

\bibitem{Raymer94}
MG~Raymer, Ml~Beck, and D~McAlister.
\newblock Complex wave-field reconstruction using phase-space tomography.
\newblock {\em Physical review letters}, 72(8):1137, 1994.

\bibitem{Home2006}
JP~Home, MJ~McDonnell, DM~Lucas, G~Imreh, BC~Keitch, DJ~Szwer, NR~Thomas,
  SC~Webster, DN~Stacey, and AM~Steane.
\newblock Deterministic entanglement and tomography of ion--spin qubits.
\newblock {\em New Journal of Physics}, 8(9):188, 2006.

\bibitem{Riebe2006}
M~Riebe, K~Kim, P~Schindler, T~Monz, PO~Schmidt, TK~K{\"o}rber, W~H{\"a}nsel,
  H~H{\"a}ffner, CF~Roos, and R~Blatt.
\newblock Process tomography of ion trap quantum gates.
\newblock {\em Physical review letters}, 97(22):220407, 2006.

\bibitem{six2016quantum}
Pierre Six, Ph~Campagne-Ibarcq, Igor Dotsenko, Alain Sarlette, Benjamin Huard,
  and Pierre Rouchon.
\newblock Quantum state tomography with noninstantaneous measurements,
  imperfections, and decoherence.
\newblock {\em Physical Review A}, 93(1):012109, 2016.

\bibitem{delaney2019quantum}
Robert Delaney, Adam Reed, Reed Andrews, and Konrad Lehnert.
\newblock Quantum state tomography of a mechanical oscillator.
\newblock {\em Bulletin of the American Physical Society}, 2019.

\bibitem{xin2017quantum}
Tao Xin, Dawei Lu, Joel Klassen, Nengkun Yu, Zhengfeng Ji, Jianxin Chen, Xian
  Ma, Guilu Long, Bei Zeng, and Raymond Laflamme.
\newblock Quantum state tomography via reduced density matrices.
\newblock {\em Physical review letters}, 118(2):020401, 2017.

\bibitem{Titchener16}
James~G. Titchener, Alexander~S. Solntsev, and Andrey~A. Sukhorukov.
\newblock Two-photon tomography using on-chip quantum walks.
\newblock {\em Optics letters}, 41(17):4079--4082, 2016.

\bibitem{titchener2018scalable}
James~G Titchener, Markus Gr{\"a}fe, Ren{\'e} Heilmann, Alexander~S Solntsev,
  Alexander Szameit, and Andrey~A Sukhorukov.
\newblock Scalable on-chip quantum state tomography.
\newblock {\em npj Quantum Information}, 4(1):19, 2018.

\bibitem{oren2017quantum}
Dikla Oren, Maor Mutzafi, Yonina~C Eldar, and Mordechai Segev.
\newblock Quantum state tomography with a single measurement setup.
\newblock {\em Optica}, 4(8):993--999, 2017.

\bibitem{PaivaSanchez}
C.~Paiva-S{\'a}nchez, E.~Burgos-Inostroza, O.~Jim{\'e}nez, and A.~Delgado.
\newblock Quantum tomography via equidistant states.
\newblock {\em Physical Review A}, 82(3):032115, 2010.

\bibitem{petz2012efficient}
D{\'e}nes Petz and L{\'a}szl{\'o} Ruppert.
\newblock Efficient quantum tomography needs complementary and symmetric
  measurements.
\newblock {\em Reports on Mathematical Physics}, 69(2):161--177, 2012.

\bibitem{petz2012optimal}
D{\'e}nes Petz and L{\'a}szl{\'o} Ruppert.
\newblock Optimal quantum-state tomography with known parameters.
\newblock {\em Journal of Physics A: Mathematical and Theoretical},
  45(8):085306, 2012.

\bibitem{rastegin2014notes}
Alexey~E Rastegin.
\newblock Notes on general sic-povms.
\newblock {\em Physica Scripta}, 89(8):085101, 2014.

\bibitem{renes2004symmetric}
Joseph~M Renes, Robin Blume-Kohout, Andrew~J Scott, and Carlton~M Caves.
\newblock Symmetric informationally complete quantum measurements.
\newblock {\em Journal of Mathematical Physics}, 45(6):2171--2180, 2004.

\bibitem{Barnett}
Stephen Barnett.
\newblock {\em Quantum information}, volume~16.
\newblock Oxford University Press, 2009.

\bibitem{beneduci2010infinite}
Roberto Beneduci.
\newblock Infinite sequences of linear functionals, positive operator-valued
  measures and naimark extension theorem.
\newblock {\em Bulletin of the London Mathematical Society}, 42(3):441--451,
  2010.

\bibitem{coecke2008povms}
Bob Coecke and {\'E}ric~Oliver Paquette.
\newblock Povms and naimark's theorem without sums.
\newblock {\em Electronic Notes in Theoretical Computer Science}, 210:15--31,
  2008.

\bibitem{beneduci2010unsharpness}
Roberto Beneduci.
\newblock Unsharpness, naimark theorem and informational equivalence of quantum
  observables.
\newblock {\em International Journal of Theoretical Physics},
  49(12):3030--3038, 2010.

\bibitem{tabia2012experimental}
Gelo Noel~M Tabia.
\newblock Experimental scheme for qubit and qutrit symmetric informationally
  complete positive operator-valued measurements using multiport devices.
\newblock {\em Physical Review A}, 86(6):062107, 2012.

\bibitem{Reck94}
Michael Reck, Anton Zeilinger, Herbert~J. Bernstein, and Philip Bertani.
\newblock Experimental realization of any discrete unitary operator.
\newblock {\em Physical Review Letters}, 73(1):58, 1994.

\bibitem{clements2016optimal}
William~R Clements, Peter~C Humphreys, Benjamin~J Metcalf, W~Steven Kolthammer,
  and Ian~A Walmsley.
\newblock Optimal design for universal multiport interferometers.
\newblock {\em Optica}, 3(12):1460--1465, 2016.

\bibitem{Rehacek}
Jaroslav {\v{R}}eh{\'a}{\v{c}}ek, Berthold-Georg Englert, and Dagomir
  Kaszlikowski.
\newblock Minimal qubit tomography.
\newblock {\em Physical Review A}, 70(5):052321, 2004.

\bibitem{Pimenta10}
M.A.D. Carvalho M.R. Barros J.G. Fonseca J. Ferraz M. Terra~Cunha W.M.~Pimenta,
  B.~Marques and S.~P{\'a}dua.
\newblock Minimal state tomography of spatial qubits using a spatial light
  modulator.
\newblock {\em Optics Express}, 18(24):24423--24433, 2010.

\bibitem{chaboyer2015tunable}
Zachary Chaboyer, Thomas Meany, LG~Helt, Michael~J Withford, and MJ~Steel.
\newblock Tunable quantum interference in a 3d integrated circuit.
\newblock {\em Scientific reports}, 5:9601, 2015.

\bibitem{meany2015laser}
Thomas Meany, Markus Gr{\"a}fe, Ren{\'e} Heilmann, Armando Perez-Leija, Simon
  Gross, Michael~J Steel, Michael~J Withford, and Alexander Szameit.
\newblock Laser written circuits for quantum photonics.
\newblock {\em Laser \& Photonics Reviews}, 9(4):363--384, 2015.

\bibitem{poulios2014quantum}
Konstantinos Poulios, Robert Keil, Daniel Fry, Jasmin~DA Meinecke, Jonathan~CF
  Matthews, Alberto Politi, Mirko Lobino, Markus Gr{\"a}fe, Matthias Heinrich,
  Stefan Nolte, et~al.
\newblock Quantum walks of correlated photon pairs in two-dimensional waveguide
  arrays.
\newblock {\em Physical review letters}, 112(14):143604, 2014.

\end{thebibliography}
\addcontentsline{toc}{chapter}{Referencias}
	
\end{document}